\documentclass[12pt,titlepage]{article}
\usepackage{amssymb}
\usepackage[dvips]{graphicx}
\usepackage{amsfonts}
\usepackage[portuguese]{babel}

\begin{document}

\title{Estados ligados em um potencial delta duplo\\
via transformada de Laplace\bigskip \\
{\small \ (Bound states in a double delta potential via Laplace transform)}}
\author{A.S. de Castro\thanks{%
E-mail: castro@pq.cnpq.br} \\
\\
Departamento de F\'{\i}sica e Qu\'{\i}mica, \\
Universidade Estadual Paulista \textquotedblleft J\'{u}lio de Mesquita
Filho\textquotedblright, \\
Guaratinguet\'{a}, SP, Brasil}
\date{}
\maketitle

\begin{abstract}
O problema de estados ligados em um potencial delta duplo \'{e} revisto com
o uso do m\'{e}todo da transformada de Laplace. Bem diferentemente de m\'{e}%
todos diretos, nenhum conhecimento acerca da descontinuidade de salto da
derivada primeira da autofun\c{c}\~{a}o \'{e} requerida para se determinar a
solu\c{c}\~{a}o.\newline
\newline
\noindent \textbf{Palavras-chave:} duplo delta, estado ligado, transformada
de Laplace.\newline
\newline
\newline

{\small \noindent The problem of bound states in a double delta potential is
revisited by means of Laplace transform method. Quite differently from
direct methods, no knowledge about the jump discontinuity of the first
derivative of the eigenfunction is required to determine the solution.
\newline
\newline
}

\end{abstract}

\section{Introdu\c{c}\~{a}o}

O uso da transformada de Laplace na equa\c{c}\~{a}o de Schr\"{o}dinger
remonta ao pr\'{o}prio Erwin Schr\"{o}dinger \cite{sch} ao lidar com o \'{a}%
tomo de hidrog\^{e}nio (veja tamb\'{e}m \cite{swa}). Mais recentemente, os
estados ligados em um potencial de Morse tamb\'{e}m foram obtidos por meio
da t\'{e}cnica da transformada de Laplace \cite{che}. A ideia subjacente ao m%
\'{e}todo da transformada de Laplace para resolver uma equa\c{c}\~{a}o
diferencial \'{e} a convers\~{a}o em uma equa\c{c}\~{a}o transformada que
possa ser resolvida com maior simplicidade. Em seguida deve-se executar a
invers\~{a}o da transformada de Laplace para obter a fun\c{c}\~{a}o original
do problema. Eis uma tarefa que pode ser \'{a}rdua e at\'{e} mesmo infact%
\'{\i}vel.

A equa\c{c}\~{a}o de Schr\"{o}dinger com um potencial constitu\'{\i}do de
uma soma de duas fun\c{c}\~{o}es delta de Dirac, doravante denominado
potencial delta duplo, tem sido usada para modelar as for\c{c}as de troca
entre os dois n\'{u}cleos no \'{\i}on de hidrog\^{e}nio molecular \cite{fro}
tanto quanto na descri\c{c}\~{a}o da transfer\^{e}ncia de um nucleon de val%
\^{e}ncia durante uma colis\~{a}o nuclear \cite{bre}. A bem da verdade, os
estados estacion\'{a}rios de uma part\'{\i}cula em um potencial delta duplo
\ ocupa as p\'{a}ginas de muitos livros-texto \cite{gas}-\cite{gri}. Os poss%
\'{\i}veis estados ligados s\~{a}o encontrados pela localiza\c{c}\~{a}o dos
polos complexos da amplitude de espalhamento ou por meio de uma solu\c{c}%
\~{a}o direta da equa\c{c}\~{a}o de Schr\"{o}dinger baseada na
descontinuidade da derivada primeira da autofun\c{c}\~{a}o, mais a
continuidade da autofun\c{c}\~{a}o e seu bom comportamento assint\'{o}tico.

Neste trabalho apresenta-se uma abordagem alternativa para busca de estados
ligados do potencial delta duplo baseada na transformada de Laplace. Com
este procedimento a equa\c{c}\~{a}o de Schr\"{o}dinger independente do tempo
transmuta-se numa equa\c{c}\~{a}o alg\'{e}brica de primeira ordem para a
transformada de Laplace da autofun\c{c}\~{a}o. O processo da invers\~{a}o da
transformada de Laplace inversa \'{e} amig\'{a}vel e a solu\c{c}\~{a}o do
problema de estados ligados n\~{a}o requer qualquer conhecimento sobre a
descontinuidade da derivada primeira da autofun\c{c}\~{a}o. A abordagem do
potencial delta duplo via transformada de Laplace, al\'{e}m de estender a
aplicabilidade do m\'{e}todo de Laplace \`{a} mec\^{a}nica qu\^{a}ntica,
fornece uma nova ponte entre o material que os estudantes tipicamente
aprendem em um curso de f\'{\i}sica matem\'{a}tica e um problema f\'{\i}sico
interessante.

\section{Os estados ligados de um potencial delta duplo}

A {transformada de Laplace}

\begin{equation}
\mathcal{L}\left\{ f\left( x\right) \right\} =\int_{0}^{\infty
}dx\,e^{-sx}f\left( x\right)  \label{l1}
\end{equation}%
de uma fun\c{c}\~{a}o de ordem exponencial, i.e. $|f\left( x\right) |\leq
Me^{\sigma x}$com $\sigma \in
\mathbb{R}
$ e $M>0$, converge se \textrm{Re}$\,s>\sigma ${\ \cite{but}}. A {%
transformada de Laplace \'{e} uma opera\c{c}\~{a}o linear e o mesmo se d\'{a}
com a transformada inversa. A propriedade de deslocamento}%
\[
\mathcal{L}\left\{ \theta \left( x-x_{0}\right) f\left( x-x_{0}\right)
\right\}
\]%
\begin{equation}
\qquad =e^{-sx_{0}}\mathcal{L}\left\{ f\left( x\right) \right\} ,\;x_{0}>0,
\label{l4}
\end{equation}%
onde%
\begin{equation}
\theta \left( x\right) =\left\{
\begin{array}{cc}
1 & {\textrm{para }}x>0, \\
&  \\
0 & \textrm{para }x<0%
\end{array}%
\right.  \label{l5}
\end{equation}%
\'{e} a fun\c{c}\~{a}o degrau de Heaviside, segue diretamente da defini\c{c}%
\~{a}o da transformada de Laplace. Tamb\'{e}m segue de (\ref{l1}){\ que }%
\begin{equation}
\mathcal{L}\left\{ {\textrm{sen\thinspace }}kx\right\} =\frac{k}{s^{2}+k^{2}}%
,\quad \mathrm{Re}\,s>0,  \label{l2}
\end{equation}%
\begin{equation}
\mathcal{L}\left\{ \cos kx\right\} =\frac{s}{s^{2}+k^{2}},\quad \mathrm{Re}%
\,s>0.  \label{l3}
\end{equation}

Usando as defini\c{c}\~{o}es%
\begin{equation}
a=\frac{2m\alpha L}{\hbar ^{2}},\quad k^{2}=\frac{2mE}{\hbar ^{2}},
\label{ab}
\end{equation}%
a equa\c{c}\~{a}o de Schr\"{o}dinger independente do tempo para uma part%
\'{\i}cula de massa $m$ sujeita a um potencial delta duplo sim\'{e}trico%
\begin{equation}
V\left( x\right) =-\alpha \left[ \delta \left( x+L\right) +\delta \left(
x-L\right) \right]  \label{pot}
\end{equation}%
pode ser escrita na forma%
\begin{equation}
\phi ^{\prime \prime }\left( x\right) +\frac{a}{L}\left[ \delta \left(
x+L\right) +\delta \left( x-L\right) \right] \phi \left( x\right) +k^{2}\phi
\left( x\right) =0,  \label{eq2}
\end{equation}%
onde a plica ($^{\prime }$) denota a derivada em rela\c{c}\~{a}o a $x$, $%
\alpha $ \'{e} uma constante real e $L>0$. Multiplicando esta equa\c{c}\~{a}%
o por $e^{-sx}$ e integrando em rela\c{c}\~{a}o a $x$ de $0$ a $\infty $:%
\[
\left( s^{2}+k^{2}\right) \mathcal{L}\left\{ \phi \left( x\right) \right\}
=s\phi \left( 0\right) +\phi ^{\prime }\left( 0\right) -\frac{a}{L}%
\,e^{-sL}\phi \left( L\right)
\]%
\begin{equation}
\qquad -\lim_{x\rightarrow \infty }e^{-sx}\left[ s\phi \left( x\right) +\phi
^{\prime }\left( x\right) \right] ,  \label{par}
\end{equation}%
onde%
\begin{equation}
\phi \left( 0\right) =\lim_{x\rightarrow 0_{+}}\phi \left( x\right) ,\quad
\phi ^{\prime }\left( 0\right) =\lim_{x\rightarrow 0_{+}}\phi ^{\prime
}\left( x\right) ,  \label{origem}
\end{equation}%
e $\mathcal{L}\left\{ \phi \left( x\right) \right\} $ \'{e} a transformada
de Laplace de $\phi \left( x\right) $. Haja vista que $\phi \left( x\right) $
e $\phi ^{\prime }\left( x\right) $ s\~{a}o limitadas no infinito, temos a
garantia da exist\^{e}ncia de $\mathcal{L}\left\{ \phi \left( x\right)
\right\} $ tanto quanto a anulabilidade da \'{u}ltima parcela de (\ref{par}%
). Resulta da\'{\i} que temos uma equa\c{c}\~{a}o alg\'{e}brica para $%
\mathcal{L}\left\{ \phi \left( x\right) \right\} $ cuja solu\c{c}\~{a}o \'{e}%
\begin{eqnarray}
\mathcal{L}\left\{ \phi \left( x\right) \right\} &=&\phi \left( 0\right) \,%
\frac{s}{s^{2}+k^{2}}  \nonumber \\
&&  \nonumber \\
&&+\,\frac{\phi ^{\prime }\left( 0\right) }{k}\frac{k}{s^{2}+k^{2}}-\frac{%
a\phi \left( L\right) }{kL}\,e^{-sL}\frac{k}{s^{2}+k^{2}}.
\end{eqnarray}%
A reconstru\c{c}\~{a}o da autofun\c{c}\~{a}o $\phi \left( x\right) $ para $%
x>0$, realizada pela invers\~{a}o da transformada de Laplace, pode ser
obtida prontamente usando (\ref{l4})-(\ref{l3}):%
\begin{eqnarray}
\phi \left( x\right) &=&\phi \left( 0\right) \cos kx+\frac{\phi ^{\prime
}\left( 0\right) }{k}\,{\textrm{sen\thinspace }}kx  \nonumber \\
&&  \nonumber \\
&&-\,\frac{a\phi \left( L\right) }{kL}\,\theta \left( x-L\right) \textrm{sen}%
\left[ k\left( x-L\right) \right] .
\end{eqnarray}%
Obviamente $\phi \left( x\right) $ n\~{a}o \'{e} quadraticamente integr\'{a}%
vel se $k\in
\mathbb{R}
$. Entanto, com uso das identidades%
\begin{eqnarray}
\mathrm{sen\,}iz &=&i\,\mathrm{senh\,}z=i\,\frac{e^{z}-e^{-z}}{2},  \nonumber
\\
&& \\
\mathrm{\cos \,}iz &=&\cosh \mathrm{\,}z=\frac{e^{z}+e^{-z}}{2},  \nonumber
\end{eqnarray}%
pode-se verificar que se $k=\pm i\xi /L$ com $\xi \in
\mathbb{R}
$ ($E<0$) e
\begin{equation}
\phi \left( L\right) =\frac{\xi \,e^{\xi }}{a}\left( \phi \left( 0\right) +%
\frac{L}{\xi }\,\phi ^{\prime }\left( 0\right) \right)  \label{qua1}
\end{equation}%
assevera-se que $\phi \left( \infty \right) =0$. Deste modo podemos escrever
a autofun\c{c}\~{a}o para estados ligados, definida no semieixo positivo $X$%
, na forma%
\begin{eqnarray}
\phi \left( x\right) &=&\phi \left( 0\right) \cosh \frac{\xi x}{L}+\phi
^{\prime }\left( 0\right) \frac{L}{\xi }\,\textrm{senh\thinspace }\frac{\xi x}{%
L}  \nonumber \\
&&  \nonumber \\
&&-\left( \phi \left( 0\right) +\frac{L}{\xi }\,\phi ^{\prime }\left(
0\right) \right)  \nonumber \\
&&  \nonumber \\
&&\times \,e^{\xi }\,\theta \left( x-L\right) \textrm{senh}\left[ \xi \left(
\frac{x}{L}-1\right) \right] .  \label{funco}
\end{eqnarray}%
N\~{a}o obstante a singularidade do potencial em $x=L$, a autofun\c{c}\~{a}o
\'{e} uma fun\c{c}\~{a}o cont\'{\i}nua. Se n\~{a}o fosse assim a equa\c{c}%
\~{a}o de Schr\"{o}dinger envolveria derivadas da fun\c{c}\~{a}o delta de
Dirac. A continuidade de $\phi \left( x\right) $ em $x=L$ implica que
\begin{equation}
\phi \left( L\right) =\phi \left( 0\right) \cosh \xi +\phi ^{\prime }\left(
0\right) \frac{L}{\xi }\,{\textrm{senh\thinspace }}\xi .
\end{equation}%
Esta \'{u}ltima rela\c{c}\~{a}o combinada com (\ref{qua1}) resulta em%
\[
\phi \left( 0\right) \left( 1-\frac{a}{\xi }\,e^{-\xi }\,\cosh \xi \right)
\]%
\begin{equation}
\qquad \qquad +\,\phi ^{\prime }\left( 0\right) \frac{L}{\xi }\left( 1-\frac{%
a}{\xi }\,e^{-\xi }\,\textrm{senh}\,\xi \right) =0.  \label{Phipil}
\end{equation}

\begin{figure}[th]
\begin{center}
\includegraphics[width=9cm, angle=0]{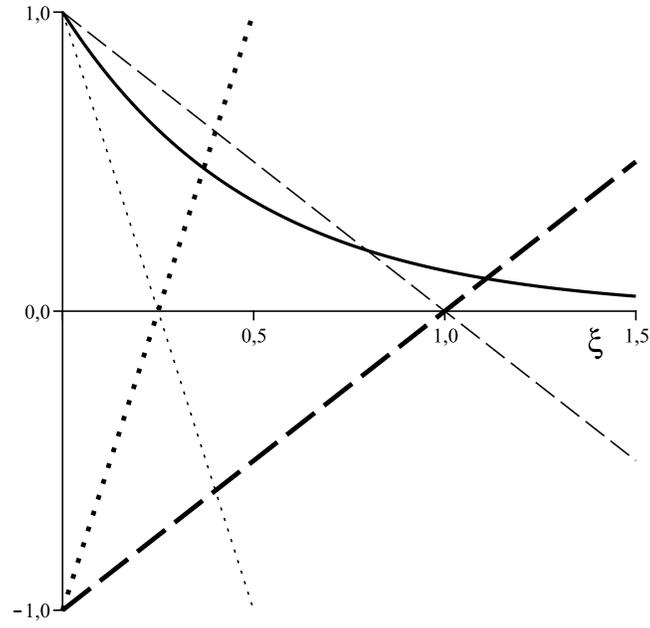}
\end{center}
\par
\vspace*{-0.1cm}
\caption{Esbo\c{c}o da condi\c{c}\~{a}o de quantiza\c{c}\~{a}o $e^{-2\protect%
\xi }=|1-2\protect\xi/a| $ para $a>0$. Curva cont\'{\i}nua para $e^{-2%
\protect\xi }$. Curvas tracejadas ($a>1$) e pontilhadas ($a<1$) para $|1-2%
\protect\xi/a|$. Curva espessa para $\protect\phi$ par, e curva delgada para
$\protect\phi$ \'{\i}mpar.}
\label{Fig1}
\end{figure}

Haja vista \ que o potencial \'{e} par sob a troca de $x$ por $-x$ (a fun%
\c{c}\~{a}o delta de Dirac \'{e} invariante sob invers\~{a}o espacial), a
extens\~{a}o da autofun\c{c}\~{a}o (\ref{funco}) para todo o eixo $X$ pode
ser expressa como uma fun\c{c}\~{a}o de paridade definida pela imposi\c{c}%
\~{a}o de condi\c{c}\~{o}es de contorno apropriadas sobre $\phi \left(
x\right) $ e $\phi ^{\prime }\left( x\right) $ na origem. Por causa da
continuidade da autofun\c{c}\~{a}o e sua derivada em $x=0$ (para $L\neq 0)$,
estas condi\c{c}\~{o}es podem ser cominadas de duas formas distintas: a fun%
\c{c}\~{a}o par obedece \`{a} condi\c{c}\~{a}o de Neumann homog\^{e}nea $%
\phi ^{\prime }\left( 0\right) =0$, enquanto a fun\c{c}\~{a}o \'{\i}mpar
obedece \`{a} condi\c{c}\~{a}o de Dirichlet homog\^{e}nea $\phi \left(
0\right) =0$. Deste modo a equa\c{c}\~{a}o (\ref{Phipil}) torna-se uma equa%
\c{c}\~{a}o para a vari\'{a}vel $\xi $. Portanto, para $\phi \left(
-x\right) =+\phi \left( x\right) $ temos%
\begin{equation}
\phi \left( x\right) =\phi \left( 0\right) \left\{
\begin{array}{cc}
\cosh \frac{\xi x}{L} & {\textrm{para }}|x|\leq L, \\
&  \\
\cosh \xi \,e^{-\xi \left( |x|/L-1\right) } & {\textrm{para }}|x|\geq L,%
\end{array}%
\right.  \label{s1}
\end{equation}%
com a condi\c{c}\~{a}o de quantiza\c{c}\~{a}o%
\begin{equation}
e^{-2\xi }=\frac{2\xi }{a}-1.  \label{quap}
\end{equation}%
Por outro lado, para $\phi \left( -x\right) =-\phi \left( x\right) $ temos%
\begin{eqnarray}
\phi \left( x\right) &=&\phi ^{\prime }\left( 0\right) \frac{L}{\xi }
\nonumber \\
&&  \nonumber \\
&&\times \left\{
\begin{array}{cc}
{\textrm{senh\thinspace }}\frac{\xi x}{L} & {\textrm{para }}|x|\leq L, \\
&  \\
{\textrm{senh\thinspace }}\xi \,\mathrm{\varepsilon }\left( x\right) \,e^{-\xi
\left( |x|/L-1\right) } & {\textrm{para }}|x|\geq L,%
\end{array}%
\right.
\end{eqnarray}%
onde $\mathrm{\varepsilon }\left( x\right) =x/|x|$ ($x\neq 0$) \'{e} a fun%
\c{c}\~{a}o sinal, e a condi\c{c}\~{a}o de quantiza\c{c}\~{a}o manifesta-se
agora na forma%
\begin{equation}
e^{-2\xi }=1-\frac{2\xi }{a}.  \label{quai}
\end{equation}%
J\'{a} que a fun\c{c}\~{a}o $e^{-2\xi }$ \'{e} limitada entre os valores $0$
e $1$ ao passo que $|1-2\xi /a|$ n\~{a}o se inclui dentro destes limites
quando $a<0$, podemos inferir que n\~{a}o h\'{a} possibilidade de solu\c{c}%
\~{a}o para estados ligados se $a<0$ (potencial repulsivo). Para um
potencial atrativo ($a>0$), a natureza do espectro resultante das solu\c{c}%
\~{o}es das equa\c{c}\~{o}es transcendentais (\ref{quap}) e (\ref{quai})
podem ser visualizadas na Figura \ref{Fig1}, onde constam esbo\c{c}os dos
membros direito e esquerdo de (\ref{quap}) e (\ref{quai}). As abscissas das
interse\c{c}\~{o}es de $e^{-2\xi }$ e $|1-2\xi /a|$ fornecem as solu\c{c}%
\~{o}es desejadas. Da\'{\i}%
\begin{equation}
E=-\frac{\hbar ^{2}\xi ^{2}}{2mL^{2}}.  \label{energia}
\end{equation}%
Pode-se depreender da Figura \ref{Fig1} que sempre h\'{a} uma e somente uma
solu\c{c}\~{a}o para o caso de uma autofun\c{c}\~{a}o sim\'{e}trica mas a
exist\^{e}ncia de uma solu\c{c}\~{a}o para o caso de uma autofun\c{c}\~{a}o
antissim\'{e}trica sucede t\~{a}o somente quando $a>1$. Isto se d\'{a}
porque $1-2\xi $ oscula $e^{-2\xi }$ em $\xi =0$. Seja l\'{a} como for, o
estado fundamental corresponde a uma autofun\c{c}\~{a}o par.

\section{Coment\'{a}rios finais}

Os leitores podem verificar que a metodologia aqui apresentada pode ser
estendida com facilidade para um potencial constitu\'{\i}do de uma soma de
um n\'{u}mero arbitr\'{a}rio de fun\c{c}\~{o}es delta de Dirac dispostas
simetricamente em rela\c{c}\~{a}o \`{a} origem. Contudo, o caso de um
potencial delta de Dirac localizado na origem requer uma modifica\c{c}\~{a}o
na defini\c{c}\~{a}o da transformada de Laplace que inclua a origem no dom%
\'{\i}nio de integra\c{c}\~{a}o. De fato,%
\begin{equation}
\mathcal{L}_{-}\left\{ f\left( x\right) \right\} =\int_{0_{-}}^{\infty
}dx\,e^{-sx}f\left( x\right)
\end{equation}%
tem sido usada por alguns autores \cite{au1}-\cite{au2} para incorporar as
condi\c{c}\~{o}es sobre $f\left( x\right) $ em $x=0_{-}$ . Entretanto, o uso
de $\mathcal{L}_{-}$ no caso de um potencial delta de Dirac localizado na
origem demanda o conhecimento da descontinuidade da derivada primeira da
autofun\c{c}\~{a}o.

\bigskip

\bigskip

\noindent{\textbf{Agradecimentos}}

O autor \'{e} grato ao CNPq pelo apoio financeiro. Um \'{a}rbitro atencioso
contribuiu para proscrever incorre\c{c}\~{o}es constantes na primeira vers%
\~{a}o deste trabalho.

\newpage

\end{document}